\documentclass[twocolumn,showpacs,preprintnumbers,amsmath,amssymb]{revtex4}
\usepackage{graphicx}
\usepackage{dcolumn}
\usepackage{bm}
\usepackage{float}
\begin{document}
                                                                                
\title{Congestion and decongestion in a communication network}
                                                                                
\author{Brajendra K. Singh}
\email{braj@phys.sinica.edu.tw}
\author{Neelima Gupte}
\email{gupte@chaos.iitm.ac.in}
\affiliation{Department of Physics, Indian Institute of Technology Madras, Chennai 600 036, India.}
\date{\today}
\begin{abstract}
We study network traffic dynamics in a  two dimensional
communication network with regular nodes and hubs. 
If the network experiences heavy message traffic, 
congestion
occurs due to finite capacity of the nodes. We discuss strategies to
manipulate hub capacity and hub connections to relieve hub congestion.
We find that the betweenness centrality (BC) criterion  is useful for
identifying hubs which are most likely to cause congestion, and that the
addition of assortative connections to  hubs of high BC
relieves congestion most efficiently.
\end{abstract}



\maketitle
The study of congestion in network dynamics is a topic of recent interest 
and practical importance \cite{Trafficnetwork}.
Telephone networks, traffic networks and
computer networks all experience serious delays in the transfer of
information due to congestion problems\cite{Jamming}. Network congestion 
occurs when too many hosts simultaneously try to send too much data through a network.
Various factors such as capacity, band-width and network topology play
an important role in contributing to traffic congestion. Optimal structures
for communication networks have been the focus of recent
studies \cite{Optstruc}. It has been established that the manipulation
of node-capacity and network 
capacity can effect drastic improvement in the performance and efficiency
of load-bearing networks \cite{Janaki}. Protocols which
can efficiently manipulate these factors to relieve congestion at high
traffic densities in communication networks can be of practical importance.  
In this paper, we study the problem of  congestion in a two dimensional
communication network of hubs and nodes with a large number of messages
travelling on the network and discuss
efficient methods by which traffic congestion can be reduced by minimal
manipulation of the  hub capacities and connections. We conclude
that the addition of assortative connections to the hubs of highest 
betweenness centrality \cite{BC} is the most effective way to 
relieve congestion problems.

We study traffic congestion for
a 
model network with local clustering developed in Ref.\cite{BrajNeel}.
This network consists of a two{-}dimensional lattice with two types of nodes,
ordinary nodes and hubs (See Fig. 1).  Each ordinary node
is connected to its nearest-neighbours, whereas
the hubs are connected to 
all nodes  within
a given  area of influence
defined as a square of side $2k$ centered around the hub\cite{FN}. 
The hubs are randomly distributed on the lattice
such that no two hubs are separated by less than a minimum distance,
$d_{min}$. Constituent nodes in the overlap areas of hubs  acquire
connections to all the hubs whose influence areas overlap.  

\begin{figure}[tbp]
\includegraphics[scale=0.35]{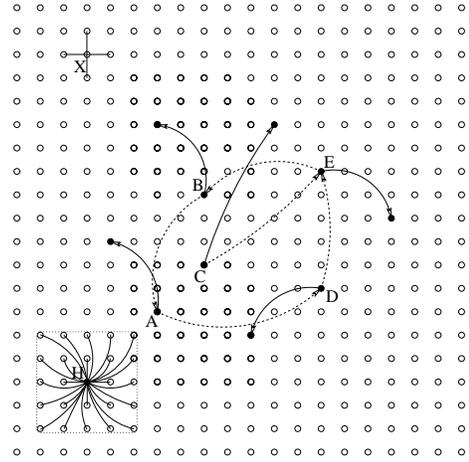}
\caption{\label{fig:epsart} A  2-d lattice with 
regular nodes with $4$ edges connected to all nearest neighbours (as explicitly shown for node $X$) and 
hubs (filled circles) connected to  all constituent nodes within their influence area (see the hub $H$). One way assortative connections between hubs (filled circles) are also shown.
Two-way connections can be visualised  by making each arrow bidirectional. The $dashed$ arrows represent the case when the assortative linkage is between any two of the top five hubs (labelled $A-E$), while the $solid$ arrows show the case when the other end point is selected randomly from the rest of the hubs.}
\end{figure}
We simulate message traffic  on this system. Any
node  can function as a source or target node for a   
message and can also be a temporary message holder or router. 
The metric distance between any pair of source 
($is,js$) and target ($it,jt$) nodes on the network is
defined to be the Manhattan distance $D_{st}=|is-it|+|js-jt|$.
The traffic flow on the
network is governed by the following rules:

Creation: A given number $N_{m}$ of source and target pairs separated by a fixed distance $D_{st}$ are
randomly selected on the lattice. 
All  source nodes start sending messages to the selected recipient
nodes simultaneously, however, each node can act as a source for
only one message during a given run. The number of source and
target pairs of a given separation $D_{st}$ are limited by the lattice
size. A phase transition between a congested state and a 
non-congested state can take place as a function of $N_{m}$ and 
$D_{st}$. These quantities are chosen to have values such that congestion 
can take place on the network, i.e. at least one message does not reach
its target during a fixed run for all the studied realisations of the
network, and source and target configurations.

Routing: It is easy to see that the shortest paths between source and target pairs on the lattice go through hubs. Hence it is optimal to route messages through hubs.
 The node which contains the message at a given time  (the current message holder $i_t$) tries to send the message towards  a temporary target, which is chosen to be a hub in its vicinity. This  hub  
(the temporary target $H_T$)
 must be the hub nearest $i_t$, and its distance from
the target must be less than the distance between $i_t$ and
the target. Once the temporary target is identified, the routing proceeds as
 follows:
i)If the $i_t$ is an ordinary node, it sends the message to its nearest neighbour towards $H_T$.
ii)If the $i_t$ is a hub, it forwards the message to one of its
constituent nodes, which is nearest to the final target.
iii)If the would-be recipient node is occupied, then the message waits
for a unit time step at  $i_t$. If the desired node is still
occupied after the waiting time is over, $i_t$ selects
any unoccupied node of its remaining neighbours and hands over the message.
In case all the remaining neighbours are occupied, the
message waits at $i_t$ until one of them is free.
iv)When a constituent node of $H_T$, receives the message, it
 sends the message directly to the hub. If $H_T$ is occupied, then the
message waits at the constituent node until the hub is free.
v)When a hub designated as $H_T$ receives a message, it sends
the message to a peripheral node in the direction of the target, which then 
chooses a new hub as the new temporary target and sends a message in its
direction.

During peak traffic, when many messages run, some of the hubs, which
are located such that many paths run through them, have to
handle more messages than they are capable of holding simultaneously.
Messages tend to jam in the vicinity of such hubs
leading to congestion in the network. Similar phenomena have been observed 
in many transportation networks \cite{Trafficnetwork, Jamming}.
It is therefore important to devise strategies which are capable of 
relieving the congestion in the network. 

If the hub capacity is  crucial in the prevention of congestion, 
can it be enhanced  to relieve congestion?
If so, which
are the hubs whose capacities should be augmented? Can decongestion be achieved in the network without major (and expensive) additions of capacity? We test out these ideas
in the current study.

A crucial quantity which identifies the hubs at which congestion occurs 
is called the `betweenness centrality' \cite{BC, Jamming} . It is useful to define a quantity, the co-efficient of  betweenness centrality (CBC), to
be the ratio of the number of messages $N_k$ which pass through a given hub $k$ to the total number of messages 
which run simultaneously  i.e. $CBC=\frac{N_k}{N}$. 
Hubs of high CBC clearly function as potential congestion points in the network. A systematic augmentation of capacity at these hubs may be useful in 
relieving the congestion in the network.
 The behaviour of many 
communication networks in real life indicates that a few hubs may be 
responsible for the worst cases of congestion, and the significant addition 
of capacity at
these hubs alone may go a long way towards relieving network congestion. 
Again, if typical separations between source and target are high, the central region 
of the lattice  is likely to  contain  hubs of high CBC. It may thus be useful to 
augment the capacity of hubs in the central region.
We compare three distinct ways of capacity enhancement which utilise the above
ideas.

In the first method (CBC$_1$), hub capacities are enhanced  in proportion to
their CBC values. The new capacity of any hub is assigned by
multiplying its CBC with a maximum capacity factor  $ \kappa$
($\kappa=2$ for our simulations) with
fractional values set to their
nearest integer number. If this assignment gives zero hub capacity to
some hub, its previous capacity is restored. 
While this method enhances the capacity of many hubs,
each hub capacity is enhanced by a very small amount.
The second way of enhancing hub capacity (CBC$_2$), viz. the significant addition of hub capacity at a few crucial hubs, is based on the selection
of $\eta$ top ranking hubs ranked according to $CBC$.  
Our simulations use $\eta=5$ and $\kappa=5$.
Lastly, using the idea that the central region (CR) of the lattice is likely to contain hubs which tend to congest,  the capacity of the hubs in this region is enhanced. Here, since the hubs are identified by their geographic location on the lattice, calculations of the CBC can be avoided.

We compare the performance of the enhancement methods outlined above for  a network of ($100\times 100$)
nodes with overlap parameter $d_{min}=1$ for hub densities upto $4.0\%$. The total number
of messages $N_m =2000$ and  $D_{st}=142$. The  length of the run is fixed at $4 D_{st}$.
The average fraction of messages which reach their destination and the average
travel time of the messages which reach are measures of the efficiency of
the network and are calculated over $1000$ configurations.
\begin{table}[h]
\caption{ This table shows $F$ the  fraction of messages delivered during a run as a function of the hub density $D$. The second  column shows $F$ for the baseline viz. the lattice with hubs of unit capacity and the remaining columns show the  fraction of messages delivered for capacity enhancement by proportional enhancement depending on $CBC$ ($CBC_1$), enhancement of top 5 $CBC$ hubs ($CBC_2$), and enhancement of  capacity in the central region (CR).}
\begin{tabular}{|c| c| c| c| c| }
\hline
  D & $F_{Base}$ & $F_{CBC_1}$ & $F_{CBC_2}$ & $F_{CR}$ \\
\hline 
  0.10&      0.06225&       0.08096&     0.18260&     0.07510\\ 
  0.50&      0.17441&       0.20744&     0.27144&     0.26875\\ 
  1.00&      0.30815&       0.34846&     0.39229&     0.48916\\
  2.00&      0.51809&       0.56974&     0.60946&     0.79287\\
  3.00&      0.68611&       0.74625&     0.77793&     0.92596\\
  4.00&      0.81786&       0.86692&     0.89181&     0.97395\\ 
\hline
\end{tabular}
\end{table}
Column 2 of Table 1 lists the fraction of messages which reach their
destination for the  hub densities in  column 1. Here,  
each hub and each node has unit capacity and thus can 
only hold a single message at a given time. A second message which arrives at 
the given hub at the same time has to wait in queue until the hub is cleared. 
The fraction of successful
transmissions goes up while the average travel time, $T_{avg}$, decreases,
as the hub density increases.
This is due to the fact that, as the hub density goes up, the
number of short paths between given sources and targets increases and the number of messages which can reach
their target within the given run  goes up because of the existence of more
alternate pathways.

Columns 3 and 4 of Table 1 list the results of the first two methods of
capacity enhancement, viz. 
$CBC_1$ and  $CBC_2$
with the top 5 hubs  enhanced. It is clear that both the enhancement
methods clear the congestion more efficiently than the base-line data,
both in terms of travel times and the number of messages which reach the destination. The $CBC_2$ method performs better than the $CBC_1$ method. 
Column 5 of Table 1 lists the results of the enhancement of capacity of the 
hubs in the central region of the lattice (of size $49 \times 49$ nodes)
to the value $\kappa= 2$ (the CR method).  The decrease in congestion is not significant below the hub density of $0.5\%$.
However, at the hub densities between $1.0\%$
and $2.5\%$ the decrease is substantially higher
than that observed in the other methods, as a large number of hubs now get
their capacities enhanced. At yet higher hub densities, the performance
of the $CR$ method saturates even though it does better than $CBC_1$ and $CBC_2$.
Unfortunately, this is a high cost method, as  a huge number of hubs need to be enhanced to get this performance at
high densities.
In contrast, the  $CBC_2$ method which only enhances five hubs performs better at
low densities. We must, however note that on an average, the $CBC_2$ method
only effects a $10 \%$ improvement over the base-line in terms of the number of messages delivered successfully to the target. 

\begin{table}[h]
\caption{ This table shows $F$ the  fraction of messages delivered during a run as a function of the hub density $D$. The second  column shows $F$ for the baseline viz. the lattice with hubs of unit capacity and the remaining columns show the  fraction of messages delivered for the assortative strategies described in the text.} 
\begin{tabular}{|c| c| c| c| c| c| }
\hline
  D & $F_{Base}$ & $F_{CBC_{2a}}$ &$F_{CBC_{2b}}$ & 
$F_{CBC_{2c}}$ & $F_{CBC_{2d}}$\\
\hline
  0.10&      0.06225&     0.41583 &   0.41220 &  0.66554 &  0.75690 \\
  0.50&      0.17441&     0.46484 &   0.47420 &  0.58882 &  0.70206 \\
  1.00&      0.30815&     0.63798 &   0.64728 &  0.72041 &  0.81193 \\
  2.00&      0.51809&     0.84177 &   0.85024 &  0.88792 &  0.92364 \\
  3.00&      0.68611&     0.94249 &   0.94678 &  0.95901 &  0.96914 \\
  4.00&      0.81786&     0.98033 &   0.98173 &  0.98536 &  0.98860 \\
\hline
\end{tabular}
\end{table}
Earlier studies on branching hierarchical networks indicate that the manipulation of capacity and connectivity  together can lead to major 
improvements in the performance and efficiency of the network \cite{Janaki}.
In addition, studies of the present  network \cite{BrajNeel}
indicate that the introduction of  a small number of assortative connection per hub  has a 
drastic effect on the travel times of messages. It is therefore interesting to investigate whether introducing 
connections between hubs of high $CBC$ has any effect on relieving congestion.

The connections can be introduced in a variety of ways. Two
possible ways (both shown in Fig. 1) are: i) One way as well as two way connections
can be introduced between the top five hubs (i.e. the five hubs with the
highest values of CBC).  ii) Assortative
connections are introduced between each the top five hubs and any one of
the remaining hubs (excluding the top five) randomly. These can be one
way or two way. The capacity of the top $5$ hubs is enhanced to $5$, so that these schemes are variants of the CBC$_2$ scheme. We note that more than one hub per connection is possible for each one of the two cases.    

Table 2 shows the results of adding assortative connections. At the lowest hub density ($0.1 \%$) the total fraction of messages delivered increases from $ 6 \%$ to $41 \%$ as soon as
one -way assortative connections are introduced between the top $5$ hubs
(see columns labelled base-line and $CBC_{2a}$).
This increases marginally if one way connections are introduced between
the top $5$ hubs and randomly chosen hubs from the remaining hubs
($CBC_{2b}$). However, the introduction of 
two-way connections between the top $5$ increases the number of messages
delivered from $6 \%$ (baseline) to  $66\%$ ($CBC_{2c}$). Setting up  two-way connections between the top $5$ hubs
and randomly chosen other hubs increased the number of messages which
were successfully delivered  to $75 \%$ ($CBC_{2d}$).
At higher hub densities, there was not much difference between the delivery efficiency of different types of assortative connections, but every type of assortative connection performed significantly better than the base-line. 
In fact, a comparison of the data sets of Table 1 and Table 2 shows that,
at any arbitrary hub density, every one of the assortative strategies performs better than all previous
strategies which enhance capacity alone. 
Thus, it is clear that the addition of assortative connections is a very efficient way of relieving congestion.

\begin{figure}[tbp]
\includegraphics[scale=0.5]{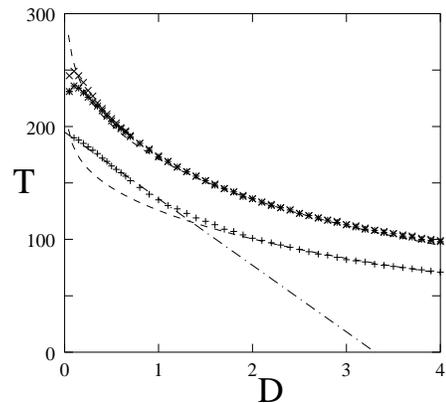}
\caption{\label{fig:epsart} The figure shows average travel times for $2000$ messages as functions of hub-density. The base-line behaviour is indicated by $asterisks$ and that on the  $CBC_2$ lattice by $crosses$ and the $CBC_{2d}$ lattices by $pluses$. The fitted lines are described in the text.}
\end{figure}
The comparison of average travel times for the messages which are
successfully delivered is also interesting. The capacity enhancement methods 
discussed earlier show hardly any change in the travel times of
messages which are successfully delivered over the average travel times
for the base-line. On the other hand, the introduction of assortative connections cuts travel times by about $20 \%$. Two way assortative connections between the top $5$ hubs and randomly chosen other hubs perform best in terms of travel times. 
The behaviour of travel times as a function of hub density is plotted in 
Fig. 2 for the case where the top 5 hubs have a single extra connection
with randomly chosen hubs other than these five, for the baseline and
the CBC$_2$ cases. The plots for the baseline as well as the CBC$_2$ cases
can be fitted by a stretched
exponential function $f_1(x)=A_1 exp(-c_1 x^{\alpha_1})(x)^{-\beta_1}$ where
$A_1=220$, $c_1=0.25$, $\alpha_1=0.77$ and $\beta_1=0.083$. The travel times
for the case of assortative connections show rather different behaviour.
At low hub densities the travel times fall off linearly and can be
fitted by the function $g(x)=-mx + C$ where $m=59$ and $C=195$. At high
hub densities a good fit can be obtained by the function
$f_2(x)=A_2 exp (-c_2 x^{\alpha_2})(x)^{- \beta_2}$, where $A_2=155$,
$c_2=0.21$, $\alpha_2=0.85$ and $\beta_2=0.08$. We note that a stretched exponential fall-off has been observed earlier for the base-line \cite{BrajNeel}. However, the cross-over to power-law behaviour seen for the case of assortative connections in Ref. \cite{BrajNeel} is not seen here, as the total number of assortative connections added here is much smaller. Instead, we have a linear fall off up to hub densities of $1\%-1.5\%$, and stretched exponential behaviour thereafter \cite{FN1}. 

\begin{figure}
\includegraphics[scale=0.7]{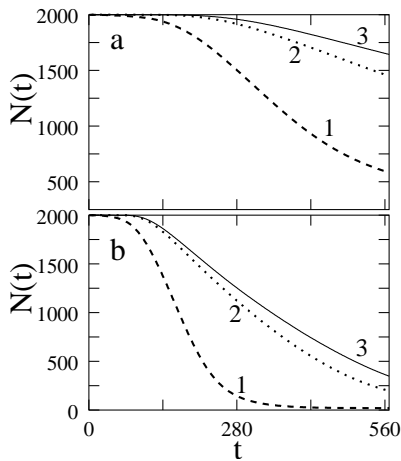} 
\caption{\label{fig:epsart} Plot of $N(t)$, the number of messages
running on the lattice as a function of $t$ at a) low hub density (50
hubs), b) high hub density (400 hubs). The curve labelled `1' shows
the behaviour on the lattice with assortative connections, the curve
labelled `2' shows that of the lattice with enhanced capacity ($CBC_2$)
and that labelled `3' shows the behaviour of the base-line.} 
\end{figure}
The quantity $N(t)$, the total number of messages running in the system
at a given time $t$, is also a useful quantifier of the efficiency of the
system in delivering messages, as the number of messages decreases 
as they are delivered to the desired target. We plot this quantity in 
Fig. 3(a) (low hub densities) and Fig. 3(b) (high hub densities) for the four cases defined above. It is clear that the
addition of  two-way connections from the top five
hubs (after capacity augmentation) to randomly chosen hubs from the
remaining hubs relieves the congestion extremely rapidly in comparison
to the base-line at both low and high hub densities.

To summarise, the addition of assortative connections to hubs of
high betweenness centrality is an extremely efficient way of relieving
congestion  in a network. While the augmentation of capacity at such
hubs also contributes towards decongestion, it does not work as
efficiently as the addition of assortative connections. Efficient
decongestion can be achieved by the addition of  extra
connection to a very small number of hubs of high betweenness
centrality. Decongestion is achieved most rapidly when two-way
connections are added from the hubs of high betweenness centrality to
other randomly chosen hubs. However, other ways of adding assortative
connections such as one way connections, or one-way and two way
connections between the hubs of high CBC also work reasonably well. 
We note that this method is a low cost method as very few  extra 
connections are added to as few as five hubs. The data indicates that 
a large augmentation of capacity would be required to achieve similar 
levels of decongestion by the addition of capacity alone.  
The methods used here are general and can be carried over to other types 
of networks as well.
We therefore
hope our methods will find useful applications in realistic situations.

We thank CSIR, India for partial support in this work.


\begin{thebibliography}{}
\bibitem{FN}
The shape of the influence area does not affect the results seen \cite{BrajNeel}.
\bibitem{FN1}
A small inflection to the linear behaviour can be seen around hub-densities of 
$0.5 \%$.
\end{thebibliography}
\end{document}